\documentclass[useAMS,usenatbib]{mn2e}

\title[General Relativistic Radiation Hydrodynamics]
{Equations of General Relativistic Radiation Hydrodynamics from Tensor
Formalism}
\author[Myeong-Gu Park]{Myeong-Gu Park$^{1}$\thanks{E-mail:
mgp@knu.ac.kr}
\\ $^{1}$Kyungpook National University, Daegu 702-701,
Korea}

\begin{document}

\date{Accepted 2004 December 15. Received 2005 February 28}

\pagerange{\pageref{firstpage}--\pageref{lastpage}} \pubyear{2002}

\maketitle

\label{firstpage}

\begin{abstract}
Radiation interacts with matter via exchange of energy and momentum.
When matter is moving with a relativistic velocity or when the
background spacetime is strongly curved, rigorous relativistic
treatment of hydrodynamics and radiative transfer is required. Here, we
derive fully general relativistic radiation hydrodynamic equations from
a covariant tensor formalism. The equations can be applied to any
three-dimensional problems and are rather starightforward to understand
compared to the comoving frame-based equations. Current approach is
applicable to any spacetime or coordinates, but in this work we
specifically choose the Schwarzschild spacetime to show explicitly how
the hydrodynamic and the radiation moment equations are derived. Some
important aspects of relativistic radiation hydrodynamics and the
difficulty with the radiation moment formalism are discussed as well.
\end{abstract}


\section{Introduction}

Radiation interacts with matter. The dynamics and energy contents of
matter are influenced by surrounding radiation while radiation itself
is concurrently determined by surrounding matter. So one needs to solve
the hydrodynamic and the radiative transfer equations simultaneously
and self-consistently.

When matter is moving with a relativistic velocity, various
relativistic effects should be taken into account: Doppler shift,
aberration, time dilation, relativistic beaming, and so forth. These
make physical quantities observer-dependent: one needs to specify
carefully the observer from whose point of view physical quantities are
measured. The same is true for matter and radiation in curved
spacetime. Gravitational redshift, time dilation, loss cone effect, and
so forth should be taken into account. To consistently deal with all
these relativistic effects, one needs to build the radiation
hydrodynamics within a fully relativistic framework.

\citet{tho30} formulated special relativistic theory of radiative
transfer to deal with radiative viscosity in a differentially moving
media. \citet{lin66} subsequently developed a covariant form of general
relativistic photon transport equation: a partial differential equation
in time, space, angle, and frequency. A straightforward solution to
such a fully angle- and frequency-dependent radiative transfer equation
would be a direct integration of the multi-dimensional, partial
differential equation. Although this can be done in principle by a
finite difference method \citep{hau91,lieb04}, characteristic methods
which reduce the transfer equation to a single ordinary differential
equation along characteristic rays are more practical and successfully
applied to special relativistic \citep{mih80} and general relativistic
problems \citep{schm78,schn88,schn89,zan96}. However, even these
characteristic methods have been limited so far to spherically
symmetric and steady cases. And if one is interested in the dynamics of
the gas flow rather than in the details of the emergent spectrum,
simpler description of the radiation field may suffice.

\citet{lin66} also derived relativistic radiation moment equations from
the transport equation. In the radiation moment formalism, a finite
number of angular moments, generated from the specific intensity, are
used to describe the radiation field. \citet{tho81} generalized the
moment formalism to projected symmetric trace-free (PSTF) tensors to
derive the relativistic version of moment equations up to an arbitrary
order. PSTF formalism has been successfully applied to a variety of
spherically symmetric problems
\citep{fla82,fla84,tur88,nob91,nob93,zam93}.

Almost all aforementioned formalisms are based on the comoving frame:
matter and radiation quantities and their directional derivatives are
defined and described within the comoving frame. It is most convenient
to define various matter and radiation quantities and describe their
interactions in the comoving frame, a frame that moves along with
matter. But, the directional derivatives in the comoving frame are
rather awkward because the velocity of matter varies in space and in
time.

One can also present the radiation hydrodynamic equations in Eulerian
framework: derivatives appear in much simpler form in the Eulerian
frame. The most straightforward Eulerian frame is a frame fixed with
respect to the central object, which we call the fixed frame. The
methodology of using comoving radiation quantities within fixed-frame
coordinates was first employed by \citet{mih80}, which we will call the
mixed-frame formalism. \citet{par93} also started from covariant tensor
equations for energy-momentum to derive the mixed-frame radiation
hydrodynamic equations for spherically symmetric systems: matter and
radiation quantities and their derivatives were in covariant forms
under fixed-frame coordinates while the interactions between radiation
and matter were described by the comoving quantities. This approach
makes the equations easier to interpret and apply to astrophysical
problems. Recently, special relativistic radiation hydrodynamic
equations for three-dimensional problems were derived by the same
approach \citep{par04}. In this work, we extend this mixed-frame
formalism to arbitrary spacetime or coordinates. We show specifically
how the three-dimensional, general relativistic radiation hydrodynamic
equations are derived in the Schwarzschild geometry. We also discuss
some aspects of relativistic radiation hydrodynamics as well as the
difficulty in closing the radiation moment equations. We put \(c=1\) in
most equations except in a few cases where \(c\) is explicitly written
for clarity.

\section[]{Tensor Equations}

\subsection{Matter}

The energy-momentum tensor of an ideal gas is
\begin{equation}\label{eq:Tgab}
T^{\alpha\beta} \equiv  \omega_g U^\alpha U^\beta + P_g
g^{\alpha\beta},
\end{equation}
where \( U^\alpha \) is the four velocity of the gas and \( \omega_g
\equiv \varepsilon_g + P_g \) the gas enthalpy per unit proper volume
which is the sum of the gas energy density \(\varepsilon_g\) and the
gas pressure \(P_g\). The functional dependence of the gas enthalpy
\(\omega_g\) on the gas temperature \(T\) can become complex,
especially in the transrelativistic regime \citep{ser86}.

\subsection{Radiation}

The energy and momentum of the radiation field are represented by the
radiation stress tensor that consists of zeroth, first, and second
moments in angle of the radiation field,
\begin{equation}\label{eq:Rab}
R^{\alpha\beta} = \int \!\! \int I(\mathbf{n},\nu) n^\alpha n^\beta
d\nu d\Omega,
\end{equation}
where  \(I(x^\alpha;{\bf n},\nu)\) is the specific intensity of photons
moving in direction \(\bf n\) on a unit sphere of projected tangent
space with the frequency \(\nu\) measured by a fiducial observer,
$p^\alpha$ the four-momentum of photons, and $n^\alpha \equiv
p^\alpha/h\nu$. Since \( I_\nu / \nu^3 \) which is proportional to the
photon distribution function and \(\nu d\nu d\Omega\) are
frame-independent scalars, \(R^{\alpha\beta}\) is a contravariant
tensor.

\subsection{Radiation Hydrodynamic Equations in Covariant Form}

Since mass and energy are equivalent in relativity, it is the particle
number density, rather than the mass density, that is conserved in
relativistic hydrodynamics. The continuity equation, therefore, should
represent the conservation of the number density,
\begin{equation}\label{eq:nU}
\left( n U^\alpha \right)_{; \alpha} = 0 .
\end{equation}

In the absence of any external force other than the radiative
interaction, the total energy-momentum tensor of gas plus radiation is
conserved,
\begin{equation}\label{eq:TRcons}
\left( T^{\alpha\beta} + R^{\alpha\beta} \right)_{; \beta} = 0.
\end{equation}

One can define the radiation four-force density to specifically
describe the energy and momentum transferred from radiation to gas
\citep{mih84},
\begin{equation}\label{eq:Gab}
G^\alpha \equiv \frac{1}{c} \int d\nu \int d\Omega [ \chi I({\bf
n},\nu) - \eta ] n^\alpha,
\end{equation}
where \(\chi\) is the opacity per unit proper length and \(\eta\) the
emissivity per proper unit volume. The combinations \(\eta/\nu^2\) and
\(\nu\chi\) are again frame-independent scalars.

This radiation four-force density is equal to the divergence of the
energy-momentum tensor for matter,
\begin{equation}\label{eq:Tab=Ga}
T^{\alpha\beta}{}_{; \beta} = G^\alpha,
\end{equation}
and to the minus the divergence of the radiation stress tensor,
\begin{equation}\label{eq:Rab=-Ga}
R^{\alpha\beta}{}_{; \beta} = -G^\alpha.
\end{equation}
When micro-physical processes for the interactions between radiation
and matter are known, equation (\ref{eq:TRcons}) can be put into two
separate equations (\ref{eq:Tab=Ga}) and (\ref{eq:Rab=-Ga}) and solved.

\section{Schwarzschild Spacetime}

The above tensor equations (\ref{eq:nU})--(\ref{eq:Rab=-Ga}) can be
explicitly written out in any spacetime or coordinates . In this work,
we specifically choose the Schwarzschild spacetime to show explicitly
how to derive more familiar hydrodynamic and radiation moment equations
and various transformation relations among radiation moments. Much
simpler, spherically symmetric case has been shown in \citet{par93}.

\subsection{Schwarzschild geometry}

The most familiar form of the Schwarzschild metric is
\begin{eqnarray}\label{eq:gab}
   d \tau^2 &=& -g_{\alpha\beta} dx^\alpha dx^\beta \nonumber \\
        &=& \Gamma^2 dt^2 - \frac{dr^2}{\Gamma^2}
           -r^2( d\theta^2 + \sin^2\theta d\phi^2 ) ,
\end{eqnarray}
where \(M\) is the mass of the central object, \( m \equiv GM/c^2 \),
and
\begin{equation}\label{eq:def_Gamma}
   \Gamma \equiv \left( 1- \frac{2m}{r} \right)^{1/2}
\end{equation}
is the lapse function.

The four-velocity of the gas, defined as
\begin{equation}\label{eq:Udef}
   U^\alpha \equiv \frac{d x^\alpha}{d\tau},
\end{equation}
satisfies the normalization condition
\begin{equation}\label{eq:U2}
   U_\alpha U^\alpha = -1,
\end{equation}
from which one can define the energy parameter
\begin{eqnarray}\label{eq:y_def}
   y &\equiv& -U_t   \nonumber \\
     &=& \left[ \Gamma^2 + (U^r)^2 + \Gamma^2 (r U^\theta)^2
   +  \Gamma^2 (r \sin\theta U^\phi)^2 \right]^{1/2}.
\end{eqnarray}

\subsection{Tetrads}

It is not trivial to define physical quantities for matter or radiation
in curved spacetime. However, one can always set up some tetrad, and
since a tetrad frame is a locally inertial frame, physical quantities
can be straightforwardly defined as in flat spacetime. Among various
tetrads, fixed and comoving tetrads are the most relevant ones. A fixed
tetrad is an orthonormal tetrad fixed with respect to the coordinates
and has base \({\bf e}_{\hat i} = \partial/\partial x^{\hat i}\) that
can be expressed in terms of coordinate base \(\partial/\partial x^i\)
as
\begin{eqnarray}
   \frac{\partial}{\partial {\hat t}}
     &=& \frac{1}{\Gamma} \frac{\partial}{\partial t} \nonumber \\
   \frac{\partial}{\partial {\hat r}}
     &=& \Gamma \frac{\partial}{\partial r} \\
   \frac{\partial}{\partial {\hat \theta}}
     &=& \frac{1}{r} \frac{\partial}{\partial \theta} \nonumber \\
   \frac{\partial}{\partial {\hat \phi}}
     &=& \frac{1}{r\sin\theta} \frac{\partial}{\partial \phi}.
     \nonumber
\end{eqnarray}
Physical quantities defined in the fixed tetrad are those measured by a
fiducial observer who is fixed with respect to the coordinates. This
observer sees matter moving with the proper velocity \(\bmath{v}\),
which is related to the four-velocity as
\begin{equation}\label{eq:vi}
   v^i = \frac{U^{\hat i}}{U^{\hat t}}
       = \frac{U_\alpha e_{\hat i}{}^\alpha}{-U_\alpha e_{\hat
       t}{}^\alpha},
\end{equation}
where \(U^{\hat i}\) and \(U^{\hat t}\) are the spatial and time parts
of the fixed tetrad components of the four-velocity and \(e_{\hat
i}{}^\alpha\) is the \(\alpha\)-th component of the tetrad base \({\bf
e}_{\hat i}\). Specifically,
\begin{eqnarray}\label{eq:v_rthphi}
   v^r      &=& \frac{1}{y} U^r, \nonumber \\
   v^\theta &=& \frac{\Gamma}{y} r U^\theta, \\
   v^\phi   &=& \frac{\Gamma}{y} r \sin\theta U^\phi. \nonumber
\end{eqnarray}
Since \(\bmath{v}\) is a three vector defined in the tetrad, \(v_i =
v^i\). We also define the Lorentz factor \(\gamma\) for \(\bmath{v}\)
as
\begin{equation}\label{eq:gamma_v}
   \gamma \equiv (1-v^2)^{-1/2} = \frac{y}{\Gamma},
\end{equation}
where \(v^2 = \bmath{v}\cdot\bmath{v} = v_i v^i = v_r^2 + v_\theta^2 +
v_\phi^2\). The value of \(v^2\) at the horizon is always 1 regardless
of \(U^i\): a fiducial observer fixed at the horizon always sees matter
radially falling in with velocity \(c\).

A comoving tetrad moves with velocity \(v^i\) relative to the fixed
tetrad and therefore is related to the fixed tetrad by the Lorentz
tranformation,
\begin{equation}\label{eq:xco2x}
   \frac{\partial}{\partial x_{co}^{\hat \alpha}}
   = \Lambda^{\hat \beta}{}_{\hat \alpha}(\bmath{v})
   \frac{\partial}{\partial x^{\hat \beta}},
\end{equation}
where \({\hat \alpha}\) and \({\hat \beta}\) denote tetrad's bases. The
components of Lorentz transformation
\(\Lambda^{\hat\alpha}{}_{\hat\beta}(\bmath{v})\) are
\begin{eqnarray}\label{eq:Lorentz}
   \Lambda^{\hat t}{}_{\hat t} &=& \gamma \nonumber \\
   \Lambda^{\hat i}{}_{\hat t} &=& \gamma v^i \\
   \Lambda^{\hat t}{}_{\hat j} &=& \gamma v_j \nonumber \\
   \Lambda^{\hat i}{}_{\hat j} &=& \delta^i{}_j + v^i v_j \frac{\gamma-1}{v^2}.
       \nonumber
\end{eqnarray}

We can also express the comoving tetrad in terms of coordinate base,
which is useful when converting the tetrad components to and from
covariant ones:
\begin{eqnarray}\label{eq:e_co}
   \frac{\partial}{\partial {\hat t_{co}}}
   &=& \frac{\gamma}{\Gamma}\frac{\partial}{\partial t}
    +  \gamma\Gamma v_r \frac{\partial}{\partial r}
    +  \gamma v_\theta \frac{1}{r} \frac{\partial}{\partial \theta}
    +  \gamma v_\phi \frac{1}{r\sin\theta} \frac{\partial}{\partial
    \phi},
       \nonumber \\
   \frac{\partial}{\partial {\hat r_{co}}}
   &=& \frac{\gamma}{\Gamma} v_r \frac{\partial}{\partial t}
    +  \Gamma \left[ 1 + (\gamma-1)\frac{v_r^2}{v^2} \right]
       \frac{\partial}{\partial r} \nonumber \\
   &+& (\gamma-1) \frac{v_r v_\theta}{v^2}\frac{1}{r} \frac{\partial}{\partial \theta}
    +  (\gamma-1) \frac{v_r v_\phi}{v^2}
       \frac{1}{r\sin\theta} \frac{\partial}{\partial \phi}, \nonumber \\
   \frac{\partial}{\partial {\hat \theta_{co}}}
   &=& \frac{\gamma}{\Gamma} v_\theta \frac{\partial}{\partial t}
    +  \Gamma(\gamma-1) \frac{v_r v_\theta}{v^2} \frac{\partial}{\partial r} \\
   &+& \left[ 1 + (\gamma-1)\frac{v_\theta^2}{v^2} \right]
       \frac{1}{r} \frac{\partial}{\partial \theta}, \nonumber \\
   &+& (\gamma-1) \frac{v_\theta v_\phi}{v^2}
       \frac{1}{r\sin\theta} \frac{\partial}{\partial \phi} \nonumber \\
   \frac{\partial}{\partial {\hat \phi_{co}}}
   &=& \frac{\gamma}{\Gamma} v_\phi \frac{\partial}{\partial t}
    +  \Gamma(\gamma-1) \frac{v_r v_\phi}{v^2} \frac{\partial}{\partial r} \nonumber \\
   &+& (\gamma-1) \frac{v_\theta v_\phi}{v^2}
       \frac{1}{r} \frac{\partial}{\partial \theta} \nonumber \\
   &+& \left[ 1 + (\gamma-1)\frac{v_\phi^2}{v^2} \right]
       \frac{1}{r\sin\theta} \frac{\partial}{\partial \phi}. \nonumber
\end{eqnarray}

Similarly, applying the inverse Lorentz transformation
\(\Lambda^{\hat\alpha}{}_{\hat\beta}(-\bmath{v})\) yields the inverse
transformation from the comoving tetrad to the coordinate base:
\begin{eqnarray}\label{eq:e_fix}
   \frac{1}{\Gamma} \frac{\partial}{\partial t}
   &=& \gamma \frac{\partial}{\partial {\hat t}_{co}}
    -  \gamma v_r \frac{\partial}{\partial {\hat r}_{co}}
    -  \gamma v_\theta \frac{\partial}{\partial {\hat \theta}_{co}}
    -  \gamma v_\phi \frac{\partial}{\partial {\hat \phi}_{co}},
       \nonumber \\
   \Gamma \frac{\partial}{\partial r}
   &=& -\gamma v_r \frac{\partial}{\partial {\hat t}_{co}}
    +  \left[ 1 + (\gamma-1)\frac{v_r^2}{v^2} \right]
       \frac{\partial}{\partial {\hat r}_{co}} \nonumber \\
   &+& (\gamma-1) \frac{v_r v_\theta}{v^2}
       \frac{\partial}{\partial {\hat \theta}_{co}}
    +  (\gamma-1) \frac{v_r v_\phi}{v^2}
       \frac{\partial}{\partial {\hat \phi}_{co}}, \nonumber \\
   \frac{1}{r} \frac{\partial}{\partial \theta}
   &=& -\gamma v_\theta \frac{\partial}{\partial {\hat t}_{co}}
    +  (\gamma-1) \frac{v_\theta v_r}{v^2}
       \frac{\partial}{\partial {\hat r}_{co}} \\
   &+& \left[ 1 + (\gamma-1)\frac{v_\theta^2}{v^2} \right]
       \frac{\partial}{\partial {\hat \theta}_{co}} \nonumber \\
   &+& (\gamma-1) \frac{v_\theta v_\phi}{v^2}
       \frac{\partial}{\partial {\hat \phi}_{co}}, \nonumber \\
   \frac{1}{r\sin\theta} \frac{\partial}{\partial \phi}
   &=& -\gamma v_\phi \frac{\partial}{\partial {\hat t}_{co}}
    +  (\gamma-1) \frac{v_\phi v_r}{v^2}
       \frac{\partial}{\partial {\hat r}_{co}} \nonumber \\
   &+& (\gamma-1) \frac{v_\phi v_\theta}{v^2}
       \frac{\partial}{\partial {\hat \theta}_{co}} \nonumber \\
   &+& \left[ 1 + (\gamma-1)\frac{v_\phi^2}{v^2} \right]
       \frac{\partial}{\partial {\hat \phi}_{co}}. \nonumber
\end{eqnarray}

\subsection{Radiation Moments}

Radiation moments can be defined from the specific intensity,
\(I_\nu(x^\alpha;{\bf n})\), either in a fixed tetrad or in a comoving
tetrad as done in flat spacetime.

The radiation energy density is equal to the zeroth moment
\begin{equation}\label{eq:E_def}
E = \int\!\!\int I_\nu d\nu d\Omega, \quad E_{co} = \int\!\!\int
I_{\nu_{co}} d\nu_{co} d\Omega_{co},
\end{equation}
where \(d\Omega\) and \(d\Omega_{co}\) are the solid angle elements in
the fixed and comoving tetrad, respectively. The radiation flux
consists of three first moments
\begin{equation}\label{eq:F_def}
F^i = \int\!\!\int I_\nu n^i d\nu d\Omega, \quad F_{co}^i =
\int\!\!\int I_{\nu_{co}} n_{co}^i d\nu_{co} d\Omega_{co},
\end{equation}
where \(i = r, \theta, \phi\). The radiation pressure tensor is
symmetric and consists of six second moments
\begin{equation}\label{eq:P_def}
P^{ij} = \int\!\!\int I_\nu n^i n^j d\nu d\Omega, \quad P_{co}^{ij} =
\int\!\!\int I_{\nu_{co}} n_{co}^i n_{co}^j d\nu_{co} d\Omega_{co}.
\end{equation}

From the definition of the radiation stress tensor (Eq. \ref{eq:Rab}),
the tetrad components of the radiation stress tensor for the fixed
tetrad are simply
\begin{equation}\label{eq:Rabfix_EFP}
R^{{\hat\alpha}{\hat\beta}} = \left(
   \begin{array}{cccc}
      E        & F^r         & F^\theta       & F^\phi \\
      F^r      & P^{rr}      & P^{r\theta}    & P^{r\phi} \\
      F^\theta & P^{r\theta} & P^{r\theta}    & P^{\theta\phi} \\
      F^\phi   & P^{r\phi}   & P^{\theta\phi} & P^{\phi\phi}
   \end{array}
       \right).
\end{equation}
For the spherically symmetric radiation field,
\(F^\theta=F^\phi=P^{r\theta}=P^{r\phi}=0\) and
\(P^{\theta\theta}=P^{\phi\phi}=(E-P^{rr})/2\). The comoving tetrad
components are similarly
\begin{equation}\label{eq:Rabco_EFP}
R_{co}^{{\hat\alpha}{\hat\beta}} = \left(
   \begin{array}{cccc}
      E_{co}        & F_{co}^r         & F_{co}^\theta       & F_{co}^\phi \\
      F_{co}^r      & P_{co}^{rr}      & P_{co}^{r\theta}    & P_{co}^{r\phi} \\
      F_{co}^\theta & P_{co}^{r\theta} & P_{co}^{r\theta}    & P_{co}^{\theta\phi} \\
      F_{co}^\phi   & P_{co}^{r\phi}   & P_{co}^{\theta\phi} & P_{co}^{\phi\phi}
   \end{array}
       \right).
\end{equation}

The contravariant components of the radiation stress tensor
\(R^{\alpha\beta}\) are related to the fixed tetrad components by the
transformation
\begin{equation}\label{eq:R2Rhat}
   R^{\alpha\beta} = \frac{\partial x^\alpha}{\partial x^{\hat\mu}}
                     \frac{\partial x^\beta}{\partial x^{\hat\nu}}
                     R^{{\hat\mu}{\hat\nu}},
\end{equation}
and contain all the curvature and coordinate specifics,
\begin{equation}\label{eq:Ralphabeta}
R^{\alpha\beta} = \left(
   \begin{array}{cccc}
      \Gamma^{-2} E                   & F^r
      & \Gamma^{-1}\frac{F^\theta}{r} & \Gamma^{-1}\frac{F^\phi}{r\sin\theta}  \\
      F^r                             & \Gamma^2 P^{rr}
      & \Gamma\frac{P^{r\theta}}{r}   & \Gamma\frac{P^{r\phi}}{r\sin\theta}  \\
      \Gamma^{-1}\frac{F^\theta}{r}   & \Gamma\frac{P^{r\theta}}{r}
      & \frac{P^{\theta\theta}}{r^2}  & \frac{P^{\theta\phi}}{r^2 \sin\theta} \\
      \Gamma^{-1}\frac{F^\phi}{r\sin\theta} & \Gamma\frac{P^{r\phi}}{r\sin\theta}
      & \frac{P^{\theta\phi}}{r^2 \sin\theta} & \frac{P^{\phi\phi}}{r^2 \sin^2\theta}
   \end{array}
       \right).
\end{equation}
%\begin{equation}\label{eq:Rab_contra}
%R^{\alpha\beta} = \left(
%   \begin{array}{cccc}
%      \Gamma^{-2}E                        & F^r
%      & \Gamma^{-1}F^\theta r^{-1}        & \Gamma^{-1} (r\sin\theta)^{-1} F^\phi  \\
%      F^r                                 & \Gamma^2 P^{rr}
%      & \Gamma r^{-1} P^{r\theta}         & \Gamma (r\sin\theta)^{-1} P^{r\phi} \\
%      \Gamma^{-1} r^{-1} F^\theta         & \Gamma r^{-1} P^{r\theta}
%      & r^{-2} P^{\theta\theta}           & r^{-2} \sin^{-1}\theta P^{\theta\phi} \\
%      \Gamma^{-1}(r\sin\theta)^{-1}F^\phi & \Gamma (r\sin\theta)^{-1} P^{r\phi}
%      & r^{-2}\sin^{-1}\theta P^{\theta\phi} & (r\sin\theta)^{-2} P^{\phi\phi}
%   \end{array}
%       \right).
%\end{equation}

Since tetrad components \(R^{\hat{\alpha}\hat{\beta}}\) and
\(R_{co}^{\hat{\alpha}\hat{\beta}}\) are Lorentz tensors, they are
related by the Lorentz boost \(\Lambda^{\hat\alpha}{}_{\hat\beta}
(\bmath{v})\),
\begin{equation}\label{eq:Rab_Rcoab}
   R_{co}^{\hat{\alpha}\hat{\beta}}
   = \Lambda^{\hat \alpha}{}_{\hat \lambda}(-\bmath{v})
     \Lambda^{\hat \beta}{}_{\hat \mu}(-\bmath{v})
     R^{\hat{\lambda}\hat{\mu}}.
\end{equation}
Substituting equations (\ref{eq:Rabfix_EFP}) and (\ref{eq:Rabco_EFP})
yields the standard transformation law between the fixed-frame
radiation moments and the comoving-frame radiation moments
\citep{mih84,mun86}, which is reproduced here for completeness:
\begin{eqnarray}\label{eq:EFPco2EFPfix}
   E_{co}      &=& \gamma^2 \left[ E - 2 v_i F^i + v_i v_j P^{ij} \right] \\
   F_{co}^i    &=& - \gamma^2 v^i E
                   + \gamma\left[ \delta_j^i + (\frac{\gamma-1}{v^2} + \gamma)
                          v^i v_j \right] F^j \nonumber \\
               & & - \gamma v_j \left[ \delta_k^i
                   + \frac{\gamma-1}{v^2} v^i v_k \right] P^{jk} \\
   P_{co}^{ij} &=& \gamma^2 v^i v^j E
                   - \gamma \left[ v^i \delta_k^j + v^j \delta_k^i
                   + 2 \frac{\gamma-1}{v^2} v^i v^j v_k \right] F^k \nonumber \\
               & & + (\delta_k^i+\frac{\gamma-1}{v^2}v^i v_k)
                     (\delta_k^j+\frac{\gamma-1}{v^2}v^j v_l) P^{kl}.
\end{eqnarray}
Since this transformation is between tetrad quantities, it is valid in
curved spacetime as well, as long as the proper velocity \(\bmath{v}\)
is used. Replacing \(\bmath{v}\) with \(-\bmath{v}\) yields the inverse
transformation. As discussed in detail in \citet{mih84}, the
comoving-frame radiation flux can be significantly different from the
fixed-frame flux due to the advection of radiation energy density and
pressure, \(v^i E + v_j P^{ij}\), and one should carefully determine
which flux is the appropriate one, especially in a lower-order
approximation.

\subsection{Radiation Four-Force Density}

The energy and momentum transfer rates, measured by a comoving
observer, are given by the comoving tetrad components of the radiation
four-force density,
\begin{equation}\label{eq:Gaco}
   G_{co}^{\hat\alpha} = \frac{1}{c} \int d\nu_{co} \int d\Omega_{co}
                    [ \chi_{co} I_{\nu_{co}} - \eta_{co} ] n_{co}^{\hat\alpha}.
\end{equation}
In terms of the usual heating function \(\Gamma_{co}\), the cooling
function \(\Lambda_{co}\) (both per unit proper volume) and the mean
opacity \(\chi_{co}\) (per unit proper length), all in the comoving
frame,
\begin{eqnarray}
   G_{co}^{\hat t} &=& \Gamma_{co} - \Lambda_{co} \label{eq:Gcot_GL}\\
   G_{co}^{\hat i} &=& \bar{\chi}_{co} F_{co}^i \label{eq:Gcoi_chi},
\end{eqnarray}
where
\begin{eqnarray}\label{eq:Gaco_GLF}
   \Gamma_{co}  &\equiv& \frac{1}{c} \int d\nu_{co} \int d\Omega_{co}
                    \chi_{co} I_{\nu_{co}}, \\
   \Lambda_{co} &\equiv& \frac{1}{c} \int d\nu_{co} \int d\Omega_{co}
                    \eta_{co}, \\
   \bar{\chi}_{co} F_{co}^i &\equiv& \frac{1}{c} \int d\nu_{co} \int d\Omega_{co}
                    \chi_{co} I_{\nu_{co}} n_{co}^i .
\end{eqnarray}
Expression in the form of equation (\ref{eq:Gcoi_chi}) is possible only
if the thermally emitted (and/or scattered) photons are isotropic in
the comoving frame so that the net momentum they remove from the gas is
zero.

The components \(G^\alpha\) are related to the comoving tetrad
components \(G_{co}^{\hat\alpha}\) by the transformation
\begin{equation}\label{eq:GalphaGa}
G^\alpha = \frac{\partial x^\alpha}{\partial x_{co}^{\hat\beta}}
           G_{co}^{\hat\beta}.
\end{equation}
Substituting \({\partial x^\alpha}/{\partial x_{co}^{\hat\beta}}\) from
equation (\ref{eq:e_co}) yields
\begin{eqnarray}\label{eq:G_Gco}
   G^t &=& \frac{\gamma}{\Gamma} \left[ G_{co}^{\hat t}
           + v_i G_{co}^{\hat i} \right]
           \nonumber \\
   G^r &=& \Gamma \left[ G_{co}^{\hat r} + \gamma v_r G_{co}^{\hat t}
           + \frac{\gamma-1}{v^2} v_r v_i G_{co}^{\hat i} \right]
           \label{eq:Gr_Gco} \\
   G^\theta &=& \frac{1}{r} \left[ G_{co}^{\hat\theta} + \gamma v_\theta G_{co}^{\hat t}
           + \frac{\gamma-1}{v^2} v_\theta v_i G_{co}^{\hat i}
           \right]
           \nonumber \\
   G^\phi &=& \frac{1}{r\sin\theta} \left[ G_{co}^{\hat\phi} + \gamma v_\phi G_{co}^{\hat t}
           + \frac{\gamma-1}{v^2} v_\phi v_i G_{co}^{\hat i} \right] .
           \nonumber
\end{eqnarray}

\section{RADIATION HYDRODYNAMIC EQUATIONS}

\subsection{Continuity Equation}

The continuity equation in the Schwarzschild cordinates (\ref{eq:gab})
is
\begin{eqnarray}\label{eq:gascontiuity}
   \frac{1}{\Gamma^2} \frac{\partial}{\partial t} (yn)
   &+& \frac{1}{r^2} \frac{\partial}{\partial r} (r^2 n U^r)
       \nonumber \\
   &+& \frac{1}{\sin\theta}  \frac{\partial}{\partial\theta}
       (\sin\theta n U^\theta)
   + \frac{\partial}{\partial\phi} (n U^\phi)
   = 0,
\end{eqnarray}
which is the conservation equation of the particle number, rather than
the mass density, because the mass density \(\rho\) in relativity
necessarily contains the internal energy that is not always conserved.

The continuity equation for spherically symmetric, steady-state flow
becomes
\begin{equation}\label{eq:acctrate}
  4\pi r^2 n U^r = - \dot{N} = {\rm constant}.
\end{equation}
One should note that the velocity \(U^r\) in equation
(\ref{eq:acctrate}) is the radial component of the four-velocity
\(U^\alpha\) which is different from the proper velocity \(v^r\). It is
the proper velocity \(v^r = y^{-1}U^r\) that is always equal to \(c\)
at the black hole horizon (\(r=2m\)) regardless of the value of \(U^r\)
while \(U^r\) is not necessarily equal to \(c\) at the horizon. This
applies to non-spherical cases as well.

\subsection{Hydrodynamic Equations}

The projection tensor \(P_\alpha{}^\beta\) projects a tensor on a
direction perpendicular to \(U^\alpha\),
\begin{equation}\label{eq:projector}
   P_\alpha{}^\beta = g_\alpha{}^\beta + U_\alpha U^\beta
       = \delta_\alpha{}^\beta + U_\alpha U^\beta .
\end{equation}
Relativistic Euler equation is obtained by projecting equation
(\ref{eq:Tab=Ga}) with \(P_\alpha{}^\beta\) to get \(P_\beta{}^\alpha
T^{\beta\lambda}{}_{;\lambda} = P_\beta{}^\alpha G^\beta\), which
becomes
\begin{equation}\label{eq:Euler_cov}
   \omega_g U^\alpha{}_{;\beta} U^\beta + g^{\alpha\beta} P_{g,\beta}
   + U^\alpha U^\beta P_{g,\beta} = G^\alpha + U^\alpha U_\beta G^\beta.
\end{equation}

Radial part of Euler equation is obtained by fixing \(\alpha = r\),
\begin{eqnarray}\label{eq:Euler_r}
  & & \omega_g U^t \frac{\partial U^r}{\partial t}
  + \omega_g U^i \frac{\partial U^r}{\partial x^i} \nonumber \\
  &+& \omega_g \frac{m}{r^2} \left[ \Gamma^2 (U^t)^2
                - \Gamma^{-2}(U^r)^2 \right] \nonumber \\
  &-& \omega_g \frac{\Gamma^2}{r} \left[ (r U^\theta)^2
                            + (r\sin\theta U^\phi)^2 \right]
                            \nonumber \\
  &+& U^r U^t \frac{\partial P_g}{\partial t}
  + \Gamma^2 \frac{\partial P_g}{\partial r}
  + U^r U^i \frac{\partial P_g}{\partial x^i} \nonumber \\
  &=& - y U^r G^t + [ 1 + \Gamma^{-2} (U^r)^2 ] G^r
  + r^2 U^r U^\theta G^\theta \nonumber \\
  & & \quad + r^2 \sin^2\theta U^r U^\phi G^\phi .
\end{eqnarray}
The term \(m r^{-2}\) on the left-hand side represents the
gravitational acceleration and \(\Gamma^2 r^{-1} [ (r U^\theta)^2 +
(r\sin\theta U^\phi)^2 ] \) the centrifugal acceleration. The
right-hand side of the equation consists of all four components of
\(G^\alpha\), not just \(G^r\). In a spherically symmetric case, the
equation becomes \citep{par93}
\begin{eqnarray}\label{eq:Euler_spherical}
  \frac{y}{\Gamma}\frac{\partial U^r}{\partial t}
  &+& \frac{1}{2}\frac{\partial (U^r)^2}{\partial r} + \frac{m}{r^2}
  + \frac{y}{\Gamma}\frac{U^r}{\omega_g}
    \frac{\partial P_g}{\partial t}
  + \frac{y^2}{\omega_g}\frac{\partial P_g}{\partial r} \nonumber \\
  &=& \frac{y}{\omega_g} G_{co}^{\hat r}
  = \frac{y}{\omega_g}{\bar \chi}_{co}F_{co}^r,
\end{eqnarray}
and the comoving-frame flux \(F_{co}^r\), rather than the fixed-frame
flux \(F^r\), is directly responsible for the radial acceleration.
Therefore, even when the net flux \(F^r\) measured by an observer at
rest with respect to the coordinate is small, gas particles can still
experience significant radiative force in the presence of significant
radiation energy density \(E\) and pressure \(P^{ij}\) in the optically
thick medium, and slow down efficiently \citep{par91}.

The \(\theta\)-part of the Euler equation is
\begin{eqnarray}\label{eq:Euler_theta}
  & & \omega_g U^t \frac{\partial U^\theta}{\partial t}
  + \omega_g U^i \frac{\partial U^\theta}{\partial x^i} \nonumber \\
  &+& 2 \omega_g \frac{1}{r}U^rU^\theta
  - \omega_g \sin\theta\cos\theta (U^\phi)^2            \nonumber \\
  &+& U^\theta U^t \frac{\partial P_g}{\partial t}
  + \frac{1}{r^2} \frac{\partial P_g}{\partial\theta}
  + U^\theta U^i \frac{\partial P_g}{\partial x^i} \nonumber \\
  &=& - y U^\theta G^t + \Gamma^{-2} U^r U^\theta G^r
  + [1 + r^2 (U^\theta)^2] G^\theta \nonumber \\
  & & \quad + r^2 \sin^2\theta U^\theta U^\phi G^\phi,
\end{eqnarray}
and the \(\phi\)-part
\begin{eqnarray}\label{eq:Euler_phi}
  & & \omega_g U^t \frac{\partial U^\phi}{\partial t}
  + \omega_g U^i \frac{\partial U^\phi}{\partial x^i} \nonumber \\
  &+& 2 \omega_g \frac{1}{r}U^rU^\phi
  + 2 \omega_g \cot\theta U^\theta U^\phi           \nonumber \\
  &+& U^\phi U^t \frac{\partial P_g}{\partial t}
  + \frac{1}{r^2\sin^2\theta} \frac{\partial P_g}{\partial\phi}
  + U^\phi U^i \frac{\partial P_g}{\partial x^i} \nonumber \\
  &=& - y U^\phi G^t + \Gamma^{-2} U^r U^\phi G^r
  + r^2 U^\theta U^\phi G^\theta \nonumber \\
  & & \quad + [1 + r^2 \sin^2\theta (U^\phi)^2] G^\phi.
\end{eqnarray}

Energy equation is more readily obtained by projecting equation
(\ref{eq:Tab=Ga}) along the four-velocity \( U_\alpha
T^{\alpha\beta}{}_{;\beta} = U_\alpha G^\alpha \),
\begin{eqnarray}\label{eq:gasenergy}
  &&- n U^t \frac{\partial}{\partial t}
            \left(\frac{\omega_g}{n}\right)
  - n U^i \frac{\partial}{\partial x^i}
          \left(\frac{\omega_g}{n}\right)
  + U^t \frac{\partial P_g}{\partial t}
  + U^i \frac{\partial P_g}{\partial x^i} \nonumber \\
  &=& - yG^t + \Gamma^{-2} U^r G^r
      + r^2 U^\theta G^\theta + r^2\sin^2\theta U^\phi G^\phi \\
  &=& - G_{co}^{\hat t} = \Lambda_{co}-\Gamma_{co}.  \nonumber
\end{eqnarray}
Again, the exchange rate of energy between matter and radiation is
expressed by heating and cooling functions measured in a comoving
frame.

\subsection{Radiation Moment Equations}

The zeroth moment equation, i.e., the radiation energy equation, is
obtained by taking the time component (\(\alpha=t\)) in equation
(\ref{eq:Rab=-Ga}) and expressing \(G^\alpha\) with the corresponding
radiation moments,
\begin{eqnarray}\label{eq:radeng}
   && \frac{1}{\Gamma^2}\frac{\partial E}{\partial t}
   + \frac{1}{\Gamma^2 r^2}\frac{\partial}{\partial r}
       (r^2 \Gamma^2 F^r) \nonumber \\
   &+& \frac{1}{\Gamma r \sin\theta}\frac{\partial}{\partial \theta}
       (\sin\theta F^\theta)
   + \frac{1}{\Gamma r \sin\theta}\frac{\partial}{\partial \phi}
       (F^\phi)   \\
   &=& -G^t = -\frac{y}{\Gamma^2}\big(\Gamma_{co}-\Lambda_{co}
              + \bar{\chi}_{co} v_i F_{co}^i \big) . \nonumber
\end{eqnarray}
In equation (\ref{eq:radeng}), both the temporal and spatial
derivatives are applied to the fixed-frame radiation moments while the
interaction between matter and radiation is described by the comoving
frame quantities. This `mixed-frame' representation simplifies the
equation and makes each term easier to understand. For example,
\(\Lambda_{co}-\Gamma_{co}\) is the energy input from the matter to the
radiation field through the usual radiative heating and cooling
processes and \(\bar{\chi}_{co} v_i F_{co}^i\) is the work done by the
photons to the matter via transfer of momentum through scattering or
absorption \citep{par91}.

When the spherically symmetric radiation field does not vary in time
nor interacts with matter, equation (\ref{eq:radeng}) becomes the
`generalized luminosity conservation' equation,
\begin{equation}\label{eq:Lconstant}
  4 \pi r^2 \Gamma^2 F^r = L_\infty,
\end{equation}
where the constant \(L_\infty\) is the luminosity measured by an
observer at infinity. A static observer at a given radius \(r\) finds
the luminosity to be a function of radius,
\begin{equation}\label{eq:Lr}
  L_r \equiv 4\pi r^2 F^r = \frac{L_\infty}{1-2m/r}.
\end{equation}
This is the well-known gravitational redshift effect: locally measured
(within the gravitational potential) fixed-frame luminosity can be
larger than the luminosity seen by an observer far away (outside the
gravitational potential). For example, the fixed-frame luminosity near
the compact object can be super-Eddington even when the luminosity
measured far away is sub-Eddington \citep{par92}.

The \(r\)-part of equation (\ref{eq:Rab=-Ga}) yields the radiation
momentum equation in \(r\)-direction,
\begin{eqnarray}\label{eq:radmom_r}
   && \frac{\partial F^r}{\partial t}
   + \Gamma^2 \frac{\partial P^{rr}}{\partial r}
   + \frac{\Gamma}{r\sin\theta}\frac{\partial}{\partial \theta}
       \left( \sin\theta P^{r\theta} \right)
   + \frac{\Gamma}{r\sin\theta}\frac{\partial P^{r\phi}}{\partial \phi}
     \nonumber \\
   &+& \frac{m}{r^2}\left(E+P^{rr}\right)
   + \frac{\Gamma^2}{r}\left(2P^{rr}-P^{\theta\theta}-P^{\phi\phi}\right)
         \\
   &=& -G^r \nonumber \\
   &=& -\Gamma\bar{\chi}_{co}F_{co}^r -\Gamma \gamma v_r (\Gamma_{co}-\Lambda_{co})
   -\Gamma\frac{\gamma-1}{v^2}v_r v_i \bar{\chi}_{co}F_{co}^i. \nonumber
\end{eqnarray}
In relativistic treatment, \(v^1\)-order term \(v_r
(\Gamma_{co}-\Lambda_{co})\) and the redshift term \((m/r)(E+P)/r\)
should be included as well as the correction factors \(\Gamma\) and
\(y\). These relativistic terms are negligible in slowly moving,
optically thin flow. However, when \(\tau v \ga 1\), they significantly
affect the radiation field as well as the dynamics of the gas flow
\citep{par90,par91,par92,par95}.

The radiation momentum equation in \(\theta\)-direction from equation
(\ref{eq:Rab=-Ga}) is
\begin{eqnarray}\label{eq:radmom_theta}
   && \frac{1}{\Gamma}\frac{\partial F^\theta}{\partial t}
   + \frac{1}{r}\frac{\partial}{\partial r}
     \left(\Gamma rP^{r\theta}\right)
   + \frac{1}{r\sin\theta}\frac{\partial}{\partial \theta}
       \left( \sin\theta P^{\theta\theta} \right) \nonumber \\
   &+& \frac{1}{r\sin\theta}\frac{\partial P^{\theta\phi}}{\partial \phi}
   + \frac{2\Gamma}{r} P^{r\theta}
   - \frac{1}{r\tan\theta} P^{\phi\phi} \\
   &=& -r G^\theta \nonumber \\
   &=& -\bar{\chi}_{co}F_{co}^\theta -\gamma v_\theta (\Gamma_{co}-\Lambda_{co})
       -\frac{\gamma-1}{v^2}v_\theta v_i \bar{\chi}_{co}F_{co}^i,
        \nonumber
\end{eqnarray}
and that in \(\phi\)-direction is
\begin{eqnarray}\label{eq:radmom_phi}
   && \frac{1}{\Gamma}\frac{\partial F^\phi}{\partial t}
   + \frac{1}{r}\frac{\partial}{\partial r}
     \left(\Gamma rP^{r\phi}\right)
   + \frac{1}{r}\frac{\partial P^{\theta\phi}}{\partial \theta}
   + \frac{1}{r\sin\theta}\frac{\partial P^{\phi\phi}}{\partial \phi}
   \nonumber \\
   &+& \frac{2\Gamma}{r} P^{r\phi}
   + \frac{2}{r\tan\theta} P^{\theta\phi} \\
   &=& -r\sin\theta G^\phi \nonumber \\
   &=& -\bar{\chi}_{co}F_{co}^\phi -\gamma v_\phi (\Gamma_{co}-\Lambda_{co})
       -\frac{\gamma-1}{v^2}v_\phi v_i \bar{\chi}_{co}F_{co}^i. \nonumber
\end{eqnarray}

In a static non-relativistic case, equations
(\ref{eq:radmom_r}--\ref{eq:radmom_phi}) reduce to the familiar
diffusion equation
\begin{equation}\label{eq:diffusion}
  F^i = -\frac{1}{\bar{\chi}_{co}} P^{ij}{}_{;j}.
\end{equation}

\subsection{Discussion}

Although reducing the angle-dependent radiative transfer equation to
the radiation moment equations lowers the dimensionality of the problem
from seven (space, time, angle, and frequency) to four (space and
time), it comes with a price: while the number of moment equations
(\ref{eq:radeng}) and (\ref{eq:radmom_r}-\ref{eq:radmom_phi}) is only
four, the number of all radiation moments to be determined, \(E\),
\(F^i\), and \(P^{ij}\), is ten. This is a generic problem with moment
equations: equations at each order always contain successively higher
order moments. Astronomers routinely employ some kind of closure
relation to close the equations, for example, the Eddington factors. In
complex radiation hydrodynamic calculations, an educated guess of the
Edddington factors as functions of the optical depth is sometimes tried
(see, e.g., \citealt{tam75}, \citealt{par90}).

But it can't be stressed enough that the correct form of the Eddington
factors or any other closure relations can only be obtained by solving
the full angle-dependent radiative transfer equation (see e.g.,
\citealt{aue70}, \citealt{hum71}, \citealt{yin95}), which has been
rarely tried for three-dimensional flows and much less for relativistic
\citep{schn89} or time-dependent flows . Even in one-dimensional
systems, the viable method, iterative variable Eddington method,
requires expensive computations and, moreover, when the flow motion is
not monotonic various angle-frequency components couple along a ray and
even posing boundary conditions become a problem \citep{mih80,nob93}.
Iteration method by \citet{schm78} seems to be the only workable
approach for the general relativistic system, but it works only for
spherically symmetric cases. Although there had been a suggestion for
the general shape of the Eddginton factors for arbitrary
three-dimensional radiation hydrodynamic systems \citep{min78}, its
validity has yet to be verified by true angle-dependent radiative
transfer calculations for a variety of cases \citep{schn89}. Needless
to say, in special cases where the specific intensity or the radiation
moments can be directly calculated, the incomplete radiation moment
equations can be eschewed altogether (e.g., \citealt{cha05}).

Other related problem with the radiation moment formalism is that the
exact nature of the radiation field can be properly described only by
the infinite number of moments. Using a finite number of moments by
terminating higher order moments or assuming some kind of closure
relation can be a good enough approximation in certain conditions, but
not necessarily so in any situation: it can lead to spurious
pathological behaviours such as a radiation shock, especially when the
velocity gradient is steep \citep{tur88, dul99}.

\section{Summary}

General relativistic, radiation hydrodynamic equations are derived
using a covariant tensor formulation. Matter and radiation quantities
are defined in the fixed and comoving tetrads and transformed to
corresponding covariant forms. The interactions between matter and
radiation are described by the radiation four-force density defined in
the comoving tetrad and transformed to the covariant form. Conservation
of energy and momentum separately for matter and radiation yield the
hydrodynamic and radiation moment equations. Current approach is
applicable to any three-dimensional systems in any spacetime or
coordinates, but in this work, we show explicitly how the equations are
derived for the Schwarzschild spacetime. Certain aspects of
relativistic radiation hydrodynamics and the fundamental limitation of
the radiation moment formalism have been discussed as well.

\section*{Acknowledgments}

We thank the anonymous referee for many insightful and invaluable
comments. This work is the result of research activities (Astrophysical
Research Center for the Structure and Evolution of the Cosmos)
supported by Korea Science \& Engineering Foundation.

%\appendix

\bsp

\label{lastpage}

\end{document}